\documentclass[a4paper,10pt,notitlepage,onecolumn,twoside]{article}

\usepackage{amsmath,amssymb} 
\usepackage[dvips]{graphicx}

\usepackage{program}

\title{A New Type of Cipher}
\author{Fabio F. G. Buono \\ buonof@cli.di.unipi.it}
\date{5-1-2012}

\begin{document}

\maketitle
\tableofcontents

\begin{abstract}

We will define a new type of cipher that doesn't use neither an easy to calcualate and hard to invert 
matematical function like RSA nor a classical mono or polyalphabetic cipher.

Our cipher will use an intrinsic and not syntactic property of a sequence. The core idea of this cipher 
is a process that is  close  to changing the measurement units of the input sequence.

\end{abstract}

\section{Introduction}

As known a cipher is an algorithm that takes as input some binary sequence and gives back a ciphered sequence as output.
This process is supposed to be simple to compute but difficult to reverse unless the key is known 
(this is known as a one way trapdoor mechanism).

The problem of most known ciphers (except the one-time-pad [2][3][7][8]),
is that their ciphertexts preserve some kind of structure which can be used
from some cryptanalyst to obtain the plaintext they originated from.

Many kind of cryptoanalisis use this strategy. Therefor we shall eliminate this problem.
To be uneffected by this pathology we shall mold our cipher aroud a intrinsect non syntactic property of the information sequences, i.e.
its measurabilty as we will explain.

We are going to define a process (the measurement) that will be something like a input sequence ``weighing``, the ciphertext spawned by our cipher isn't a simple encoding in the classical meaning, but an allocation function of its value, called from now $\nabla$.

I'm going to explain in this article, some possible implementations of this idea. These implementations will help on clarifying how we use $\nabla$ in our process.

\section{The Idea}

The property we shall be using in our cipher, is a redefinition of a measure of the input sequence. 
To give an idea of what we mean by a measure of sequence we start with a simple example.
The following base two sequence \nopagebreak

\[ 1 0 1 0 0 0 0 0 1 \]

is written base ten, giving exponents in reverse sequence, as shown in the following example:

\[ 2^0 + 2^2 + 2^8 = 261 . \]

Now we define $\nabla$ as a set of positive integer such that every element of set will be smaller than previous and such that last element will be 1.

We call measure by nabla, a set of integer values. The first one is the result of integer division between decimal value of input sequence and the value of the first element of $\nabla$. 
Next value of the set is the result of integer division between decimal value of the previous element of set ( first one in this istance ) multiplied by the value of the element of $\nabla$ previously used, and so on.

Now, we will be able to define a first cipher pseudo-code.

\section{The cipher}

To show a first example of how the cipher works, we can propose the pseudocode below, there we take in input a binary sequence and translate it in its decimal value. after that we take in input an array of integral value as $\nabla$ and encode it. The output will be an array of measures by $\nabla$.
\begin{program}
\mbox{Function Cript:}
\BEGIN \\ \\ %
\rcomment{/* plaintext input as decimal value */}
Sequence \ = \ (int) \ |Input| \ (BinarySequence); \\ \\
\rcomment{/* input array of $\nabla$ */}
Nabla[] \  =  \ |Input| \ (MeasureUnit);  \\ \\               
|Output|[ \ |leght|(Nabla) \ ]; \\ \\
Index \ = \ 0; \\ \\
 
 |Forall|  \ Value  \ |in| \ Nabla[ \ ](     \\ \\
  Output[ \ Index \ ] \ =  \ Sequence \ /  \ Value;\\  \\
  Sequence = \ Sequence \ - \ ((Sequence \ Mod\ Value)* Value); \\ \\
  Index++;\\ \\
  )\\ 
Return \ Output[ \ ];
\END
\end{program}
Here is the decoding function using $\nabla$
\begin{program}
\mbox{Function Decript:}
\BEGIN \\ \\ %
InputSequence[ \ Nabla \ ] \ = \ (int) \ |Input|(BinarySequence);  \\ \\
Nabla[ \ ] \ = \ |Input|(MeasureUnit);                 \\ \\
|Output|[leght(Nabla)];                          \\ \\
Index = 0;                      \\ \\
 
|Forall| \ Value \ in \ Nabla[ \ ](  \\                    \\
  Output[ \ Index \ ] \ =  \ InputSequence[ \ nabla \ ] \ * \ Value;  \\ \\
  Index++;\\ \\
 )\\ 
Return  \ Output[ \ ];               
\END
\end{program}

This algorithm is simply to force, but there is a way to emprove this econding.

\section{Emprove the security}

To emprove the security of this cipher we change something; Let now $\nabla$, DaltaQ and DeltaR three simple set of $N$ random integer, and let $P_0$ a random integer; Now we can explain a new cipher.

\begin{program}
\mbox{Function Cript:}
\BEGIN \\ \\ %
\rcomment{/* plaintext input as decimal value */}
Sequence \ = \ (int) \ |Input| \ (BinarySequence); \\ \\
\rcomment{/* input array of $\nabla$ */}
Int \ Nabla[] \  =  \ |Input| \ (Set of integer);  \\ \\               

\rcomment{/* input array of $DeltaQ$ */}
Int \ DeltaQ[] \  =  \ |Input| \ (Set of integer);  \\ \\               

\rcomment{/* input array of $DeltaR$ */}
Int \ DeltaR[] \  =  \ |Input| \ (Set of integer);  \\ \\               

\rcomment{/* input $P_0$ */}
Int \ P0 \  =  \ |Input| \ (Integer);  \\ \\

\rcomment{/* Output array with leght of Nabla */}
|Output|[ \ |leght|(Nabla) \ ]; \\ \\

Int \ Index \ = \ 0; \\
Int \ Tmp \ = \ 0;
Int \ Value \ = \ 0; \\ \\
 
Sequence \ = \  Sequence \ * \ P0; \\ \\

|for| \ ( \ Index \ = \ 0; \ Index \ < \ lenght( \ Nabla[ \ ]; \ Index++ \ ) \ ( \\
\ Tmp \ = \ ( \ Sequence \ / \ Nabla[ \ Index \ ] \ ) \ * \ DeltaQ[ \ Index \ ];
\ Output \ [ \ Index \ ] \ = \ Tmp;
\ Sequence \ = \ ( \ Sequnce \ Mod \ Nabla[ \ Index \ ] \ ) \ * \ DeltaR[ \ Index \ ];
\ Index++;
) \\ \\

Return \ Output[ \ ];

\END
\end{program}

How we can see, our algorithm is a simple arrangement of divisions and subtractions. So it's really easy to calculate, even if impossible to come back to original input. Difficulty doesn't matter about complexity, but it comes from the possibility to find a potentially infinite set of significative results by using wrong $\nabla$, DaltaQ, DeltaR and $P_0$ values.

\section{Conclusions}

We introduced, in a very informal manner, the notion of measurement by an allocation scheme and three processes to cipher a sequence. It can be proved that both ciphers are secure, as long as the key is used only once.

It's important to say that, we can cipher a ciphered text with a new $\nabla$ improving security.

In next article [1] we going to explain a communication protocol to use this cipher for secure transfer of data, using only one security key.
We will show also mathematical proof of endurance of this cipher.

See you soon and thanks.

Special thanks to Doctor Luigi Borasi and Doctor Alessandro Bompadre and Ing. Nicola Papazafiropulos.

\section{Reference}

\noindent
[1] Entropy Measure Comunications, to be published.

[2] C.E. Shannon. Communication theory of secrecy systems. Bell System Technical Journal, 1949

[3] H. Feistel. Cryptography and Computer Privacy. Scientific American, Vol. 228, No. 5, 15 (1973)

[4] U. M. Maurer. A simplified and Generalized Treatment of Luby-Rackoff Pseudorandom Permutation
Generators. EuroCrypt '92, Springer LNCS v.658, pp.239-255, 1992

[5] M. Luby, C. Rackoff. How to construct pseudorandom permutations from pseudorandom functions. SIAM
Journal on Computing, Vol. 17, No. 2, pp 373-376, 1988

[6] M. Naor, O. Reingold, On the construction of pseudo-random permutations: Luby-Rackoff revisited,
Journal of Cryptology, Vol. 12, pp. 29-66, 1999

[7] Nicolas T. Courtois, Josef Pieprzyk. Cryptanalysis of Block Ciphers with Overdefined Systems of
Equations http://eprint.iacr.org/2002/044.pdf, 2002

[8] A. Biryukov, D. Khovratovich, Related-key Cryptanalysis of the Full AES-192 and AES-256,
https://cryptolux.org/mediawiki/uploads/1/1a/Aes-192-256.pdf, 2009

\end{document}